\documentclass[11pt]{article}

\usepackage{amsmath,amssymb,amsfonts}
\usepackage{graphicx}
\usepackage{hyperref}
\newcommand{\myfig}[3]{ 
\begin{figure}
\centering
\includegraphics[width=#2cm]{#1}\caption{#3}\label{#1}
\end{figure}
}

\begin{document}

\def\Tr{{\rm Tr }\ }

\newcommand{\ack}[1]{[{\bf Pfft!: {#1}}]}
\newcommand\bC{\mathbb{C}}
\newcommand\bP{\mathbb{P}}
\newcommand\bR{\mathbb{R}}
\newcommand\bZ{\mathbb{Z}}

\newcommand{\Ccal}{{\cal C}}

\newcommand{\Xtil}{X'}

\newcommand\dlangle{\langle\langle}
\newcommand\drangle{\rangle\rangle}
\newcommand{\bra}{\langle}
\newcommand{\ket}{\rangle}
\newcommand{\Rmath}{\mathbb{R}}
\newcommand{\Zmath}{\mathbb{Z}}

\newcommand{\beq}{\begin{equation}}
\newcommand{\eeq}{\end{equation}}
\newcommand{\beqn}{\begin{eqnarray}}
\newcommand{\eeqn}{\end{eqnarray}}

\newcommand{\beql}[1]{\begin{equation}\label{eq:#1}}
\newcommand{\eq}[1]{(\ref{eq:#1})}
\newcommand{\eref}[1]{(\ref{#1})}

\newcommand{\pa}{\partial}
\newcommand{\mlambda}{{\underline{\underline{\lambda}}}}

\def\str{{str}}
\def\lstr{\ell_\str}
\def\gstr{g_\str}
\def\Mstr{M_\str}
\def\eps{\epsilon}
\def\varep{\varepsilon}
\def\del{\nabla}
\def\grad{\nabla}
\def\tr{\hbox{tr}}
\def\perp{\bot}
\def\half{\frac{1}{2}}
\def\p{\partial}

\renewcommand{\thepage}{\arabic{page}}
\setcounter{page}{1}

\rightline{hep-th/0602083} \rightline{ILL-(TH)-06-02} \rightline{HIP-2006-03/TH} \rightline{YITP-06-07}

\vskip 0.75 cm
\renewcommand{\thefootnote}{\fnsymbol{footnote}}
\centerline{\Large \bf Fractional S-branes on a Spacetime Orbifold}
 \vskip 0.75 cm

\centerline{{\bf Shinsuke Kawai${}^{1}$\footnote{skawai@yukawa.kyoto-u.ac.jp},
Esko Keski-Vakkuri${}^{2}$\footnote{esko.keski-vakkuri@helsinki.fi},
}}
\centerline{{\bf Robert G. Leigh${}^{3}$\footnote{rgleigh@uiuc.edu} and
Sean Nowling${}^{3}$\footnote{nowling@students.uiuc.edu}
}}
\vskip .5cm
\centerline{${}^1$\it YITP, Kyoto University, Kyoto 606-8502, Japan}
\vskip .5cm
\centerline{${}^2$\it Helsinki Institute of Physics and Department of Physical Sciences,}
\centerline{\it P.O. Box 64, FIN-00014 University of Helsinki, Finland}
\vskip .5cm
\centerline{${}^3$\it Department of Physics,
University of Illinois at Urbana-Champaign}
\centerline{\it 1110 West Green Street, Urbana, IL 61801-3080, USA}
\vskip .5cm

\setcounter{footnote}{0}
\renewcommand{\thefootnote}{\arabic{footnote}}

\begin{abstract}
Unstable D-branes are central objects in string theory, and exist also in
time-dependent backgrounds.
In this paper we take first
steps to studying brane decay in spacetime orbifolds. As a concrete
model we focus on the $\bR^{1,d}/\bZ_2$ orbifold. We point out that on a
spacetime orbifold there exist two kinds of S-branes, {\em
fractional} S-branes in addition to the usual ones. We investigate
their construction in the open string and closed string boundary
state approach. As an
application of these constructions, we consider a scenario where an
unstable brane nucleates at the origin of time of a spacetime, its
initial energy then converting into energy flux in the form of
closed strings. The dual  open string description allows for a
well-defined description of this process even if it originates at a
singular origin of the spacetime.

\end{abstract}

\newpage
\section{Introduction}\label{intro}

There has been considerable interest in constructing time-dependent
string backgrounds for cosmological model-building purposes. This is
a difficult problem in general. An easier route is to take a known
string background and alter its global structure by identifying
points under the action of a discrete group so as to generate an
interesting time-dependent background. This is the idea behind
Lorentzian orbifold constructions. The prototype cosmological toy
background is the Misner space (see \cite{Durin:2005ix} for a review),
which contains regions corresponding
to a big crunch and a big bang. The most important problem in this
type of background is to develop a resolution of the cosmological
singularity. In the case of Euclidean orbifolds, this is a known
story. An interesting part in the theory of the resolved
singularities is played by D-branes. On an orbifold, there are two
kinds of D-branes, bulk branes away from the fixed points and
fractional ones passing through the fixed points. After the
resolution, the fractional D-branes lift up to regular branes
wrapping around cycles in the resolved geometry.

It is interesting to ask how to construct D-branes in Lorentzian
orbifolds. There have been studies of e.g.
D-branes in Misner space\footnote{D-branes in other time-dependent
backgrounds have been investigated in
\cite{Hashimoto:2002nr,Alishahiha:2002bk,Dolan:2002px,Cai:2002sv,Okuyama:2002pc,Hikida:2005vd}.}
\cite{Hikida:2005ec,Hikida:2005xa}.
Here we would like to address a further issue. The D-branes of
bosonic string theory are unstable, and such branes (or
configurations of stable branes that destabilize at a subcritical
separation) exist also in supersymmetric theories. Hence one must be
able to describe their decay in Lorentzian orbifolds. Understanding
of the brane decay by itself is an important topic for completeness
of string theory. Further such processes play an important role in
many cosmological scenarios, adding to the interest in this problem
with applications in mind.

There are many ways to investigate brane decay. The most
straightforward and hence in a sense a most illuminating one is to
work on the level of worldsheet string theory and attempt to deform
the action by an operator that describes the decay in the form of a
rolling tachyon background \cite{Sen:2002nu,Sen:2002in}.
Such deformations must be exactly
marginal, i.e. preserve the conformal invariance even at large
deformations in order not to spoil the unitarity of the theory. The
construction of such exactly marginal deformations is difficult in
general, and a hurdle for studies of brane decay in Lorentzian
orbifolds.

In this paper we take the first steps to investigating brane decay
(S-branes \cite{Gutperle:2002ai}) in time-dependent string backgrounds, in particular on
spacetime orbifolds. On orbifold backgrounds, it turns out that there
exists a new class of S-branes that we call {\em fractional}
S-branes, in analogy to the fractional D-branes on Euclidean
orbifolds. As a particular example, we focus on orbifolds
$\bR^{1,d}/\bZ_2$ \cite{Balasubramanian:2002ry} where the problem of finding the exactly marginal
deformation generating the rolling tachyon background is simple.

There is another motivation for considering this background, as we have discussed in \cite{Kawai:2005jx}. So far, studies of
cosmological backgrounds in string theory and string cosmology have largely focused on models where the past history of the
Universe has been extended beyond the Big Bang into an era where it undergoes a Big Crunch with the hope that conversion into the
Big Bang is possible through a resolution mechanism to be discovered, as sketched in Figure \ref{BBS1}(a), so that the arrow of
time can be continued across. \myfig{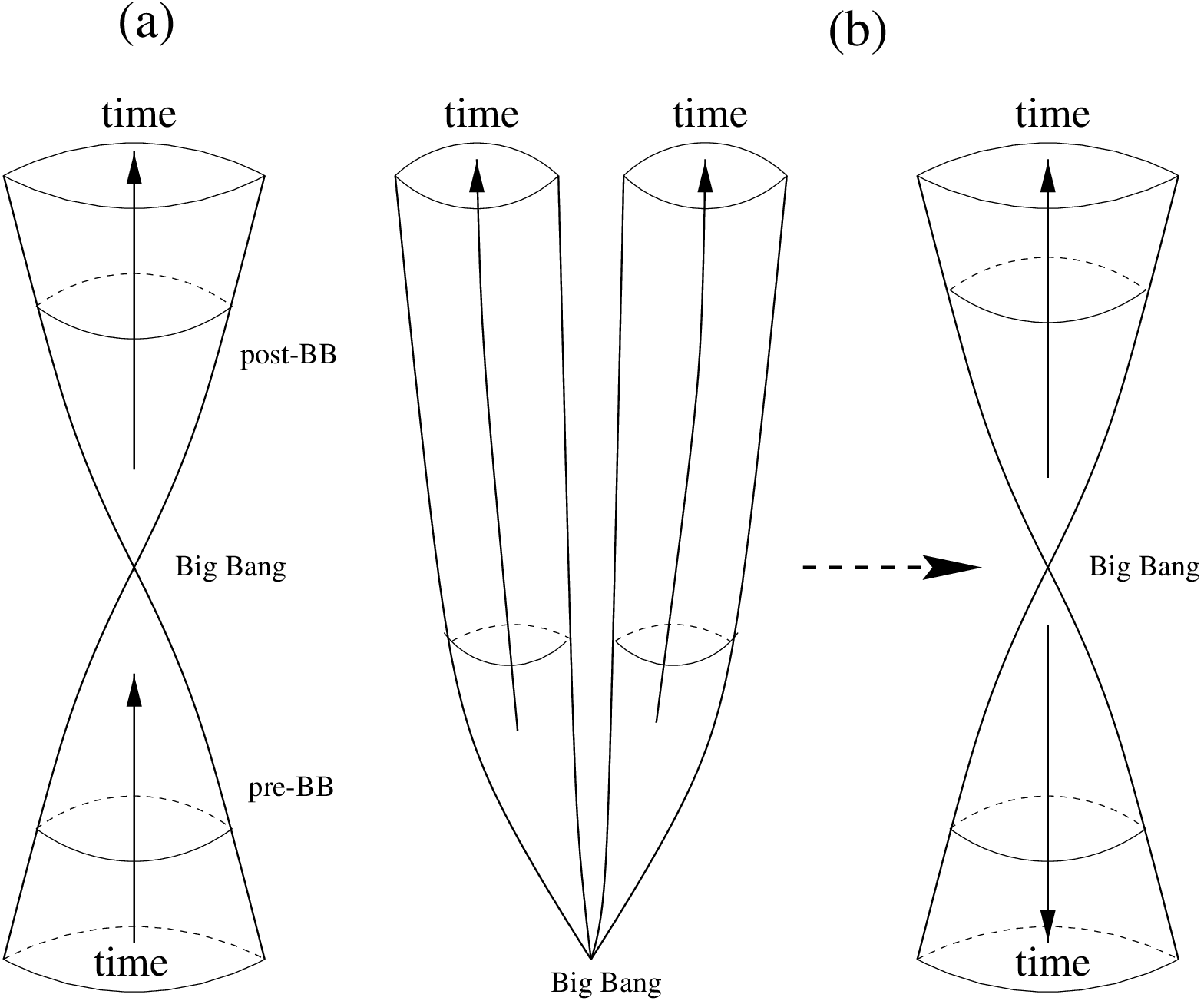}{8}{The pre-Big Bang scenario (a), and the creation of two-branched Universe from the
Big Bang, which is interpreted as a spacetime $\bZ_2$ orbifold (b).} Since the time's arrow classically `breaks' at the
singularity, one could ask if it could be taken to point to multiple directions from it. In other words, once could imagine the
Universe to be created from the Big Bang with multiple branches, each with its own arrow of time, an example is depicted in Figure
\ref{BBS1}(b). An additional ingredient in such speculations could be the role of the {\it CPT} invariance. Why does it exist in
the first place, and could it mean that Universe has two branches, each a {\it CPT} reflection of one another? This question has
been studied before in the elliptic interpretation of the de Sitter space
\cite{Schrodinger,Gibbons:1986dd,Folacci:1986gr,Parikh:2002py,Parikh:2004wh}. It might be that allowing the Universe to have
multiple branches allows for novel mechanisms for singularity resolution.

In models of branched Universes, the complication is the lack of
global time-orientability. However, it is possible that this problem
can be circumvented by a proper calculational prescription. Anyway,
in each branch, long after the Big Bang, local time-orientability is
again restored and local observers should not be affected by global
issues. A simple model of such branching of time is the orbifold
$\bR^{1,d}/\bZ_2$. After the time function has been defined on the
fundamental domain, lifting it up to the covering space can lead to
a model where time's arrow points in opposite directions away from
the initial $X^0=0$ origin. This was studied in \cite{Biswas:2003ku}.


This paper is constructed as follows. In Section \ref{orbifold} we recall the basic construction of the Lorentzian orbifold. In
Section \ref{fractional} we review briefly the construction of fractional D-branes in Euclidean orbifolds, focussing in particular
to the relation of their description in the open string sigma model and the closed string boundary state formalism. The relation
is established by considering the annulus open string partition function and its interpretation as tree-level propagation between
closed string boundary states. In Section \ref{bdycft} we then review the results of \cite{cftus} for bosonic boundary CFT and its
rolling tachyon deformations corresponding to the various S-branes. In Section \ref{bdystate} we will then collect the information
obtained in previous sections and extract out the associated deformed fractional boundary states, and use them to calculate
overlaps between the vacua in the twisted and untwisted sectors. In Section \ref{Lorentzian} we then address the Wick rotation
back to the Lorentzian  orbifold, and compute the production of closed strings in the decay and overlaps with the untwisted and
twisted vacua. Section \ref{outlook} contains conclusions and outlook.

\section{The orbifold $\bR^{1,d}/\bZ_2$}\label{orbifold}

This section collects some basic facts of the orbifold $\bR^{1,d}/\bZ_2$. We begin
with the covering space $\bR^{1,d}$ where $d$ denotes the number of spacelike
directions;
in principle we can consider any value $d\geq 0$, and a critical string background would include an additional CFT which we will not mention further.
We let $\bZ_2$ act as a {\it PT} reflection
\begin{equation}
\label{storbifold}
 (X^0,\vec{X}) \sim (-X^0,-\vec{X})
\end{equation}
on the timelike and spacelike coordinates. The origin is a fixed point of the orbifold action. By a Wick rotation
to Euclidean signature, the orbifold is related to the standard Euclidean orbifold
$\bR^{1+d}/\bZ_2$. String theory (bosonic and type II superstrings) and quantum field theory on the
orbifold (\ref{storbifold}) was investigated in \cite{Balasubramanian:2002ry} and \cite{Biswas:2003ku}. It was found
that
\begin{enumerate}
\item There are no physical states in the twisted sector in the bosonic theory
when $d>15$; in type II theory no physical states occur in the NS sector when $d>3$;
in the R sector, the twisted vacuum is the only physical state, for any $d$.

\item Negative norm states are absent at tree level in the untwisted sector.
In the twisted sector, in the bosonic theory, they are absent
when $d\geq 8$, while for $15\geq d\geq 8$ the vacuum is the only physical state
in the twisted sector. In type II theory, the twisted NS vacuum has a non-negative
norm when $d\leq 3$.

\item In the superstring theory, the one-loop partition function vanishes when
$d=3$.

\item There are no forward oriented closed causal curves when the notion of a time function has been properly defined.

\item The Hilbert space needs to be doubled to contain a backward in time propagating
image for every forward in time propagating quantum on the covering
space, the two are identified on the fundamental domain. (More detail below.)

\item The one-loop vacuum expectation value of the stress tensor vanishes almost
everywhere on the orbifold, both in string theory and in quantum field
theory. This signals the absence of any dangerous backreaction. In quantum field theory,
the only non-zero contribution is a divergence at the initial time slice. This is
related to issues regarding the resolution of the initial singular slice.
\end{enumerate}
Perhaps the most interesting feature of this orbifold is a subtlety involving
the definition of the time function. There are three different natural choices for
time orientation on the quotient, each giving rise to physically inequivalent
spacetimes (Fig. 2).
\myfig{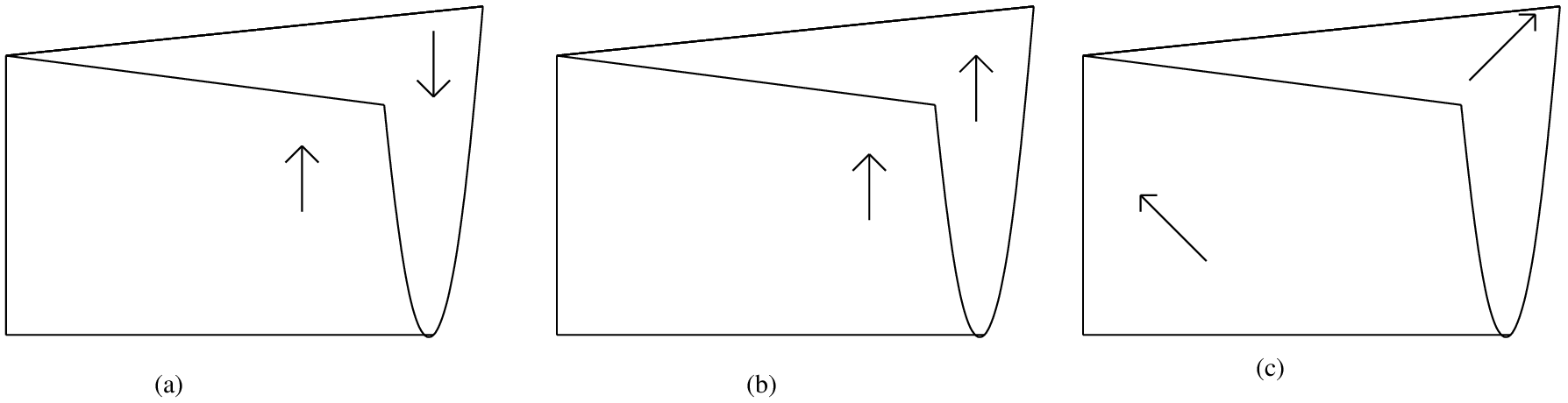}{12}{Three possible time-arrows on the quotient $\bR^{1,1}/\bZ_2$.}
In this paper we are considering the
choice (b), where time runs upwards from the X-axis with the origin representing
a ``Big Bang". On the covering space this corresponds to time axis pointing back-to-back
to opposite directions from the X-axis (Fig. 3).
\myfig{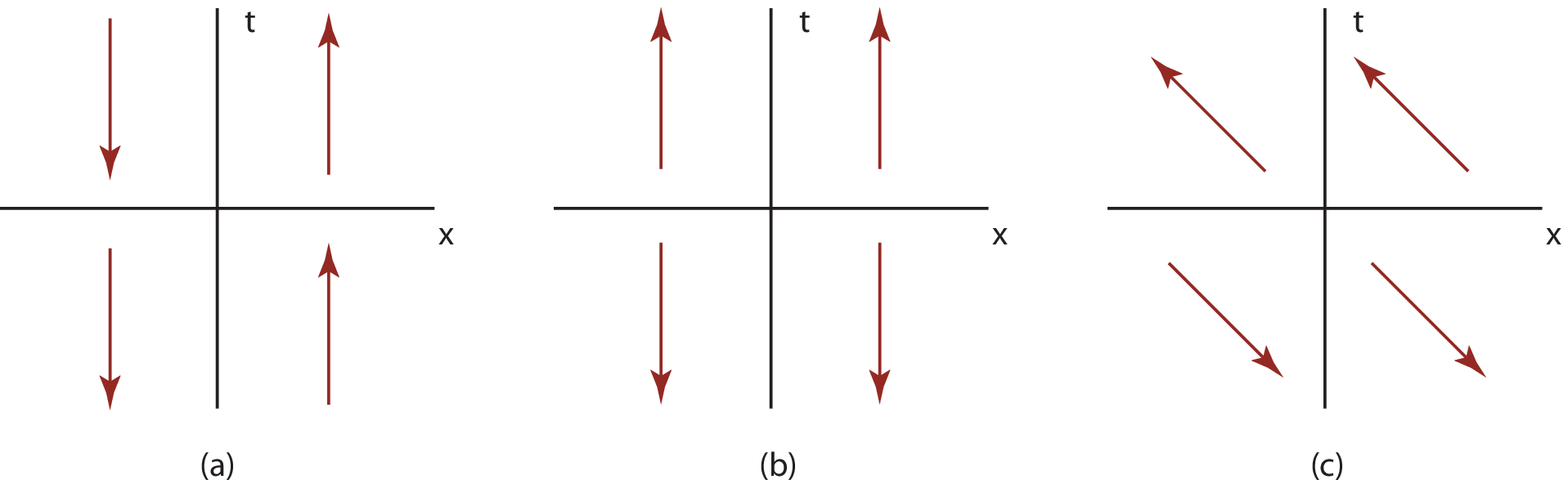}{12}{Three possible time-arrows, shown in the covering space $\bR^{1,1}$.
The time orientation line bundle is not orientable in the quotient; consequently, a time function
possesses zeroes along some locus; there are three distinct choices.}
This leads to the other
main difference with Euclidean orbifold theories. Since the $\bZ_2$ identification also
involves time reflection, forward and backward propagating quanta are identified.
Thus, when we consider string theory or quantum field theory
on the orbifold, we must start with a doubled free particle Fock space involving sectors with both sign choices of energy,
\begin{eqnarray}
 {\cal H}^+ &=& \prod_{\vec{k}} {\cal H}^+_{\vec{k}}  \ \ \ \ {\rm with}\ k^0=\omega_{\vec{k}}>0 \nonumber \\
 {\cal H}^- &=& \prod_{\vec{k}} {\cal H}^-_{\vec{k}}  \ \ \ \ {\rm with}\ k^0=-\omega_{\vec{k}}<0
\end{eqnarray}
so the full Fock space is the direct sum
\beq\label{eq:bigfock}
 {\cal H} = {\cal H}^+ \oplus {\cal H}^- \ .
\eeq
The $\bZ_2$ action then acts as an isomorphism $\cal{H}^\pm \rightarrow \cal{H}^\mp$ and
the invariant Fock space is\footnote{Note that in ordinary quantum theory (with a globally defined
time orientation), the physical Hilbert space may also be thought of as a projection of (\ref{eq:bigfock}),
where ${\cal H}^-$ is projected out.}
\beq
  {\cal H}_{inv} = {\cal H}/\bZ_2 \ .
\eeq
For example, 1-particle states have the form
\beq
  |\omega_{\vec{k}},\vec{k}\ket_{inv} = \frac{1}{\sqrt{2}}
  \left( \begin{array}{c}
           |+\omega_{\vec{k}},+\vec{k}\ket \\
           |-\omega_{\vec{k}},-\vec{k}\ket
         \end{array} \right) \ .
\eeq
In this paper we will specifically consider on-shell states in closed string theory. Since
the lower half of the covering space is the mirror image of the upper half, it is convenient
to introduce a notation $\Xtil = (\Xtil^0,\Xtil^i)=(-X^0,-X^i)$ for
the vertex operators while using standard notation for the momentum $k$,
such that $k^0>0$ for non-tachyonic states. Thus,
for example, in bosonic string theory the lowest invariant states are
\begin{equation}
  |V_T\ket_{inv} = \frac{1}{\sqrt{2}}
  \left( \begin{array}{c}
           e^{ik\cdot X}|0\ket \\
           e^{ik\cdot \Xtil}|0\ket
         \end{array} \right) \ ; \
  |V^{\mu\nu}\ket_{inv} = \frac{1}{\sqrt{2}}
  \left( \begin{array}{c}
           \partial X^\mu\bar{\partial}X^\nu e^{ik\cdot X}|0\ket \\
           \partial \Xtil^\mu \bar{\partial}\Xtil^\nu e^{ik\cdot \Xtil}|0\ket
         \end{array} \right) \ ,
\end{equation}
with the on-shell condition $M^2=-k^2=-8/l^2_s,\ 0$, respectively. We will return to this formalism later in the paper, in Section \ref{Lorentzian}.


\section{Fractional branes in Euclidean orbifolds}
\label{fractional}

The novel feature of Euclidean orbifold theories is the existence of
fractional branes, localized at the orbifold fixed points. In this
section we review the role of Chan-Paton indices in the construction
of branes in the non-compact Euclidean $\bR^D/\bZ_2$ orbifold models
(with a single fixed point). The route we will take is to deduce the
fractional brane boundary states from the open string partition
function with Chan-Paton indices.
This formalism will be carried over later to the case of deformed
boundary states, and to Lorentzian signature.

Begin with a D-brane which is pointlike in the directions of the
orbifold. An open string in the covering space then sees two
D-branes, at $X$ and $-X$.
%
%
Consequently there are 4 types of open strings which are labeled by
the branes upon which they end, summarized by the
Chan-Paton matrix
\beq \lambda =\left(%
\begin{array}{cc}
  D0-D0 & D0-D0' \\
  D0'-D0 & D0'-D0' \\
\end{array}\right).
\eeq
The $\mathbb{Z}_2$ action exchanges the
brane and image brane, in the above basis the group elements
are represented by  
\beqn
\gamma (e)=\left(\begin{array}{cc} 1&0\\0&1\\
\end{array}\right) \ , \ \gamma (g)=\left(\begin{array}{cc} 0&1\\1&0\\
\end{array}\right).\eeqn
At the fixed point, the representation
is reducible, and it is useful to work in a different basis where
\beqn \gamma (e)=\left(\begin{array}{cc} 1&0\\0&1\\
\end{array}\right) \ , \ \gamma (g)=\left(\begin{array}{cc} 1&0\\0&-1\\
\end{array}\right).\eeqn
Fractional branes are associated with the irreducible one-dimensional
representations. In the closed string language, they are described
by boundary states which we denote by $|D,\alpha\ket_{frac}$, with
$\alpha =\pm$.

Now consider the open string partition function, corresponding to an annulus diagram. Take the open string to be suspended between
two fractional branes at the same fixed point, with labels $\alpha ,\beta$. This is encoded in the partition function by inserting
appropriate projection operators $P_\alpha, P_\beta$ for each boundary into the Chan-Paton traces. They impose the projection into
the two irreducible representations, and are given by
\beqn P_+ =\left(\begin{array}{cc} 1&0\\0&0\\
\end{array}\right)\ , \ P_-=\left(\begin{array}{cc} 0&0\\0&1\\
\end{array}\right).\eeqn
The projection operators are then inserted into the
open string partition function as follows:
\beqn\label{Zab}
Z_{DD}^{\alpha,\beta} &=&
\frac{1}{2}\left([\tr~P_\alpha\gamma(e)][\tr~P_\beta\gamma(e)^{-1}]~\Tr e^{-\beta H}
+ [\tr~P_\alpha\gamma(g)][\tr~P_\beta\gamma(g)^{-1}]~\Tr ge^{-\beta H}\right)\nonumber \\
&=&\frac{1}{2}\left( \Tr e^{-\beta H} + \epsilon_{\alpha\beta}
\Tr ge^{-\beta H}\right) \ ,
\eeqn
with $\epsilon_{\pm,\pm}=1$ and $\epsilon_{\pm,\mp}=-1$.
Cardy's condition then relates the one-loop open string partition function
to the tree level closed string exchange between fractional brane boundary
states,
\beq
Z_{DD}^{\alpha,\beta}\to
\mbox{}_{frac}\langle D,\alpha |\Delta(\tilde q)|D,\beta\rangle_{frac} \ .
\eeq
It is natural to isolate the untwisted
$|\cdot\rangle_{U}$ and
twisted $|\cdot\rangle_{T}$ sector boundary states, normalized as
\beq
\mbox{}_U\langle D|\Delta|D\rangle_U = Tr\ e^{-\beta H},\ \ \ \
\mbox{}_T\langle D|\Delta|D\rangle_T = Tr\ ge^{-\beta H} \ .
\eeq
{}From (\ref{Zab}), we then read off the fractional boundary
states as linear combinations
\beq
|D,\alpha\rangle_{frac}=A_\alpha|D\rangle_U
 +B_\alpha|D\rangle_T
\eeq
with the coefficients
\beq
A_\pm=\frac{1}{\sqrt{2}},\ \ \ \
B_{\pm}=\pm\frac{1}{\sqrt{2}} \ .
\eeq
In other words, the two fractional states are
\beq
|D,\pm\rangle=\frac{1}{\sqrt{2}}\left(|D\rangle_U \pm |D\rangle_T
\right) \ .
\eeq

The regular representation is the direct sum of irreducible
representations, associated to $P_++P_-$.
Thus, the regular D-brane is identified with
$|D_o\rangle=|D,+\rangle_{frac}+|D,-\rangle_{frac}
=\sqrt{2}|D\rangle_U$.
One can check that this correctly accounts for the factors
of two and such that occur in the analysis away from the fixed
points.

The calculations in the following sections will actually involve deformations away from
Neumann-Neumann amplitude.  However, because varying
the strength of the deformation parameter smoothly interpolates between Neumann and Dirichlet boundary conditions, the Chan-Paton
structures can be treated in essentially the same manner as discussed above.


\section{Summary of CFT Results\label{bdycft}}
 We now carefully summarize the results of the companion paper \cite{cftus}. Depending on the choice of radius, there are
different computational techniques available and different results.

\subsection{The Free Theory}

First we recall that in free open string theory on a circle of radius $R$,
the Neumann-Neumann annulus amplitude has the form \cite{Oshikawa:1996dj}
\beq \label{ANNWil} {\cal A}_{NN}(\Delta\theta) =\frac{1}{\eta(q)}\sum_n q^{\alpha' (n/R+\Delta\theta/2\pi R)^2} \ . \eeq where we
have allowed for a Wilson line $\Delta\theta$. Here, $q=e^{-\pi t}$ and this has been written in the open string channel. By
Poisson resummation this becomes, with $\tilde q=e^{-2\pi/t}$,
\beq\label{eq:clchanNN}
{\cal A}_{NN} = \frac{R}{\sqrt{2\alpha'}\eta(\tilde q^2)}\sum_{m\in\bZ}
(\tilde q^2)^{m^2R^2/4\alpha'} e^{-im\Delta\theta} \ .
\eeq
This result may be factorized on the lowest lying closed string states \cite{Cardy:1989ir}
\beq
{\cal A}_{NN} (\Delta\theta) \equiv \langle N,\theta | \Delta(\tilde q)|N,\theta+\Delta\theta\rangle
\eeq
where $\Delta(\tilde q)$ is the closed string propagator.
We may then deduce the Neumann boundary state in oscillator form,
\beq
|N,\theta\rangle= 2^{-1/4} e^{\sum_k\alpha_k\tilde\alpha_k}|0\rangle_{Fock}
\otimes \sum_{m\in\bZ} e^{im\theta}|\frac{mR}{\alpha'},-\frac{mR}{\alpha'}\rangle \ .
\eeq
It has zero momentum, and is at fixed $\tilde X\equiv X_L-X_R$. Applying T-duality, we may obtain the Dirichlet boundary state at
dual radius $\tilde{R}=\alpha'/R$, \beq |D,x\rangle = 2^{-1/4}e^{-\sum_k\alpha_k\tilde\alpha_k}|0\rangle_{Fock}
\otimes\sum_{n\in\bZ} e^{-inx_0/\tilde R}|\frac{n}{\tilde R},\frac{n}{\tilde R}\rangle \eeq for a D-brane located at point $x_0$
in the dual circle.

At infinite radius, the Wilson line of the Neumann state is effectively projected to zero, while in the Dirichlet state, the momentum becomes continuous.

At self-dual radius, $R=\sqrt{\alpha'}$, the conformal dimensions are square integers, and
the spectrum can be classified by an $\widehat{SU(2)}$ current algebra (see e.g. \cite{DiFrancesco:1997nk}).
To make it explicit, (\ref{eq:clchanNN}) can be rewritten in the form \cite{Gaberdiel:2001zq}
\beq
{\cal A}_{NN} = \frac{1}{\sqrt{2}} \sum_{j=0,1/2,1,\ldots}
\chi_{j^2}^{Vir}(\tilde q^2)\chi_j^{SU(2)}(e^{-2i\Delta\theta J_0^3})
\eeq
with $SU(2)$ characters
\beq
\chi_j^{SU(2)}(g)=Tr_{j}{\cal D}^{(j)}(g) \ ,
\eeq
where ${\cal D}^{(j)}(g)$ is the matrix representing the $SU(2)$ element $g$ in representation $j$,
and Virasoro characters
\beq
\chi_{j^2}^{Vir}(\tilde q^2)=\frac{\tilde q^{2j^2}-\tilde q^{2(j+1)^2}}{\eta(\tilde q^2)} \ .
\eeq

Finally, one often uses Ishibashi states,
with the normalization
\beq\label{ishinorm}
\dlangle j,m,n | \Delta (\tilde q) |j',m',n'\drangle
=\chi_{j^2}^{Vir}(\tilde q^2)\delta_{jj'}\delta_{mm'}\delta_{nn'}
\eeq
to express the boundary state explicitly in the $\widehat{SU(2)}$ basis
\beq\label{standardexp}
|N,\theta\rangle = 2^{-1/4}\sum_{j=0,1/2,1,\ldots}\sum_{m,n=-j}^j
{\cal D}^{(j)}_{m,n}(e^{-2i\theta J_0^3})|j,-m,n\drangle \ .
\eeq

\subsection{The Free Orbifold Theory}

The $\bZ_2$ orbifold is implemented in the open string sector, apart from Chan-Paton factors, by a projection operator $\frac12 (1+g)$. The first '1' term is proportional to the results of the last subsection, and gives rise to untwisted boundary states. The $g$ term will give rise to twisted boundary states. Note that at finite radius, there are two fixed orbifold points at $x=0$ and $x=\pi R$; correspondingly, there are two discrete Wilson lines at $\theta=0,\pi$ that are fixed by the orbifold.

At self-dual radius, we find
\begin{eqnarray}
Z_{g;NN}&\equiv& Tr\ gq^{L_0-1/24}\\
&=&\frac{1}{\eta(q)}\sum_{n\in\bZ} (-1)^n q^{n^2}\\
&=&\frac{1}{\sqrt{2}}\frac{1}{\eta(\tilde q^2)}\sum_{m\in\bZ}(\tilde q^2)^{(m-1/2)^2/4} \ .
\end{eqnarray}
Writing a boundary state for the twisted states only is complicated by the presence of two fixed points. In a later section, we
will show how the lowest lying twisted modes contribute to the boundary states.

It is interesting to note that, at the self-dual radius, the orbifold partition function is T-dual to
an unorbifolded partition
function at twice the self-dual radius \cite{Ginsparg:1987eb}.  In making this equivalence we exchange the $J^3$ current at
twice the self-dual radius
with the $J^1$ current of the self-dual radius theory. Consider again the orbifold partition function
\begin{equation}
Z_{NN}= \half~Tr \ (1+g)q^{L_0-1/24} = \frac{1}{\eta(q)}\sum_{n\in\bZ} q^{ n^2} \left( \frac{1+(-1)^n}{2}\right) \ .
\end{equation}
Clearly, $n$ must be even, and we can re-write this as
\begin{equation}
Z_{NN} =
\frac{1}{\eta(q)}\sum_{n\in\bZ} q^{4 n^2}
\end{equation}
which indeed is the partition function at radius $R=\sqrt{\alpha'}/2$.
After rewriting it in the closed string channel, and
T-dualizing to radius $\tilde{R}=2\sqrt{\alpha'}$, we find the Dirichlet boundary state
with zero modes
\begin{equation}
|D,x\rangle\sim\sum_{m\in\bZ} e^{-imx_0/2\sqrt{\alpha'}}| \frac{m}{2\alpha'},\frac{m}{2\alpha'}\rangle \ .
\end{equation}
For different discrete values of $x_0$, we have boundary states which correspond to fractional brane states in the orbifold theory.
This implies a relationship between fractional branes in the self-dual radius theory,
and D-branes at twice the self-dual radius \cite{Recknagel:1998ih}.
We will elaborate this in section 4.3.1.
If we allow for the possibility of branes centered at differing positions in the twice self-dual theory, we find
\beq\label{doubleZ} Z^{DD}_{R=2R_{s.d.}} = \frac{1}{\eta(q)}\sum_{n\in\bZ}q^{(2n+\Delta x_0/2\pi \sqrt{\alpha'})^2} \ . \eeq

\subsection{The Deformed Theory}

Now we consider the boundary perturbation
%

%


\beq S_{\lambda}=\int_{\partial\Sigma} ds\ \left[ \lambda_+e^{X^0(s)/\sqrt{\alpha'}} +\lambda_-e^{-X^0(s)/\sqrt{\alpha'}} \right]
\eeq Classically (using the correlators of the undeformed theory), this perturbation is marginal, that is $h_{cl}=1$. For
$\lambda_\pm=(\lambda/2)e^{\pm X^0_0/\sqrt{\alpha'}}$, this is related to the ``full S-brane" \cite{Gutperle:2002ai}, while for
$\lambda_-=0$, we have the ``half S-brane" \cite{Larsen:2002wc}. The full S-brane corresponds to a process where a carefully
fine-tuned initial closed string configuration time evolves to form an unstable D-brane which then decays to a final state of
closed strings only. The whole process is centered around the time $X^0_0$ and is time reflection invariant about it, as evident
from writing the deformation in the form \beq\label{fullS} S_{\lambda}= \lambda\int_{\partial\Sigma} ds~\cosh
[(X^0(s)-X^0_0)/\sqrt{\alpha'}] \ , \eeq in particular the initial state of closed strings is a time reflection image of the final
state. The deformation was known to be exactly marginal by Wick rotation. Wick rotating $X^0=iX$, it becomes \beq\label{fullSE}
S_{\lambda}= -\lambda \int_{\partial\Sigma} ds~\cos[(X(s)-X_0)/\sqrt{\alpha'}] \ , \eeq which is a known exactly marginal
deformation. In practise, computations in the background (\ref{fullS}) are first performed in the Euclidean signature with
(\ref{fullSE}), and the results are then Wick rotated back to the Lorentzian signature.

One could absorb the parameter $X_0$ into the definition of the origin of time. However, for a worldsheet with multiple
boundaries, there can be a distinct deformation for each boundary component. For example, if we consider the annulus, we will
consider a boundary deformation of the form \beq\label{bdydef} S_{int}=-\lambda\int_{\partial\Sigma_1} ds
\cos\left(\frac{X-X^{(1)}_0}{\sqrt{\alpha'}}\right) -\tilde\lambda\int_{\partial\Sigma_2} ds
\cos\left(\frac{X-X^{(2)}_0}{\sqrt{\alpha'}}\right) \eeq where $\partial\Sigma_j$ are the boundary components. This is essentially
a Chan-Paton structure. Indeed in the presence of multiple branes, $\lambda$ and $\tilde\lambda$ would be replaced by matrices,
and the annulus would include overall traces for each boundary component.  For simplicity, we will assume that the $\lambda$
deformations are diagonal.  A priori, there is no need to take the cosines to be centred at the same point on different
boundaries, and the difference cannot be absorbed to the choice of the time origin.

\subsubsection{The Orbifold Case} In the orbifold $\bR^{1,d}/\bZ_2$ the $\bZ_2$ acts
by $(X^0,X^1,\ldots ,X^d)\rightarrow
-(X^0,X^1,\ldots ,X^d)$. After Wick rotation $X^0=iX$, we obtain an Euclidean orbifold $\bR^{d+1}/\bZ_2$,
where $\bZ_2$ acts by
$(X,X^1,\ldots ,X^d)\rightarrow -(X,X^1,\ldots,X^d)$. The full S-brane deformation is invariant under
the orbifold
identifications, if we choose it to be centered around $X^0=0$. In the Euclidean signature, for worldsheets
with multiple
boundaries, if allow for distinct deformations at each boundary component $\partial\Sigma_j$, we would then
need each of them to
be centered around $X=0$ (i.e., set $X^{(j)}_0=0$, but the associated parameters $\lambda_j$ can be
independent of one another).
Wick rotation back to Lorentzian signature is subtle, because of the issues with the branching of
time's arrow. This will be
discussed in Section \ref{Lorentzian}.

\paragraph{The Infinite Radius} Our basic  task, performed in \cite{cftus}, is to compute the annulus partition function for the deformed
orbifold theory. The techniques available to us depend on the choice of radius. In the non-compact case, the deformation (without
the orbifold) was studied in \cite{Polchinski:1994my} (see also \cite{Kristjansson:2004ny}).
The orbifold insertion was considered in \cite{cftus}.  An infinite radius
requires the use of fermionization techniques.  The boundary deformation is then written as a sum of fermionic bi-linears.  The
resulting action is quadratic in the fermion fields.  The non-compact partition functions are:

\beq \label{eq39}Z_1 = \frac{1}{\eta(q)}\int_0^1 d\zeta_-\sum_{m\in\bZ} q^{(m+\alpha(\zeta_-))^2}
\eeq
and
\beq
 Z_{g}=\frac{1}{\eta(q)}\sum_{\zeta_-=0,1/2}\sum_{m\in\bZ}(-1)^{m}q^{(m+\alpha(\zeta_-))^2} \ .
\eeq
Here,
\beqn \label{eq39}\sin\pi \alpha =\left(
\sin^2\left(\frac{\pi}{2}(\lambda-\tilde{\lambda})\right)\cos^2\left(\pi\zeta_-\right)
+\cos^2\left(\frac{\pi}{2}(\lambda+
\tilde{\lambda})\right)\sin^2\left(\pi\zeta_-\right)\right)^{1/2} .
\eeqn
In both expressions, $\zeta_-$ corresponds to the
fractional part of the open string's momenta.

\paragraph{The Finite Radius} If one is interested in the deformed orbifold theory
at finite radius one can construct that theory, in the particular case of a rational radius,
by implementing a suitable projection
(a translation orbifold).  At self-dual radius one may also use the adsorption technique because
the boundary deformation is
proportional to an $\widehat{su(2)}$ generator.  The function $Z_1$ was computed at self-dual radius
in \cite{Callan:1994ub}.
In \cite{cftus}, this result was reviewed and clarified and the function $Z_g$ was computed.
The self-dual results are:
\beq
Z_{1;\lambda,\tilde{\lambda}} = \frac{1}{\eta(q)}\sum_{n\in\bZ} q^{(n+(\lambda-\tilde\lambda)/2)^2}
\eeq
\beq
Z_{g;\lambda,\tilde{\lambda}} = \frac{1}{\eta(q)}\sum_{n\in\bZ}(-)^n q^{(n+(\lambda-\tilde\lambda)/2)^2} \ .
\eeq
Note that if we add
these results we get a partition function of a theory on a circle of twice the self-dual radius,
\beq
Z_{\lambda,\tilde{\lambda}} = \frac{1}{\eta(q)}\sum_{n\in\bZ} q^{(2n+\Delta \lambda /2)^2} \ ,
\eeq
with $\Delta \lambda = \lambda -\tilde\lambda$.
This can be identified with the previous partition function (\ref{doubleZ})
for D-branes in the twice self-dual radius circle theory, with the identification
\beq
   \Delta x_0 = \Delta \lambda \pi \sqrt{\alpha'} \ .
\eeq The parameter $x_0$ plays the role of the center-of-mass coordinate of the D-brane in the $2R_{s.d.}$ theory, whereas the
parameter $\lambda$ is associated with varying the open string boundary conditions in the $R_{s.d.}$ orbifold theory. The
relationship between deforming through the fractional branes in the $R_{s.d.}$ orbifold theory and the D-branes in the $2R_{s.d.}$
theory is shown in Fig. 4, elaborating upon \cite{Recknagel:1998ih}. \myfig{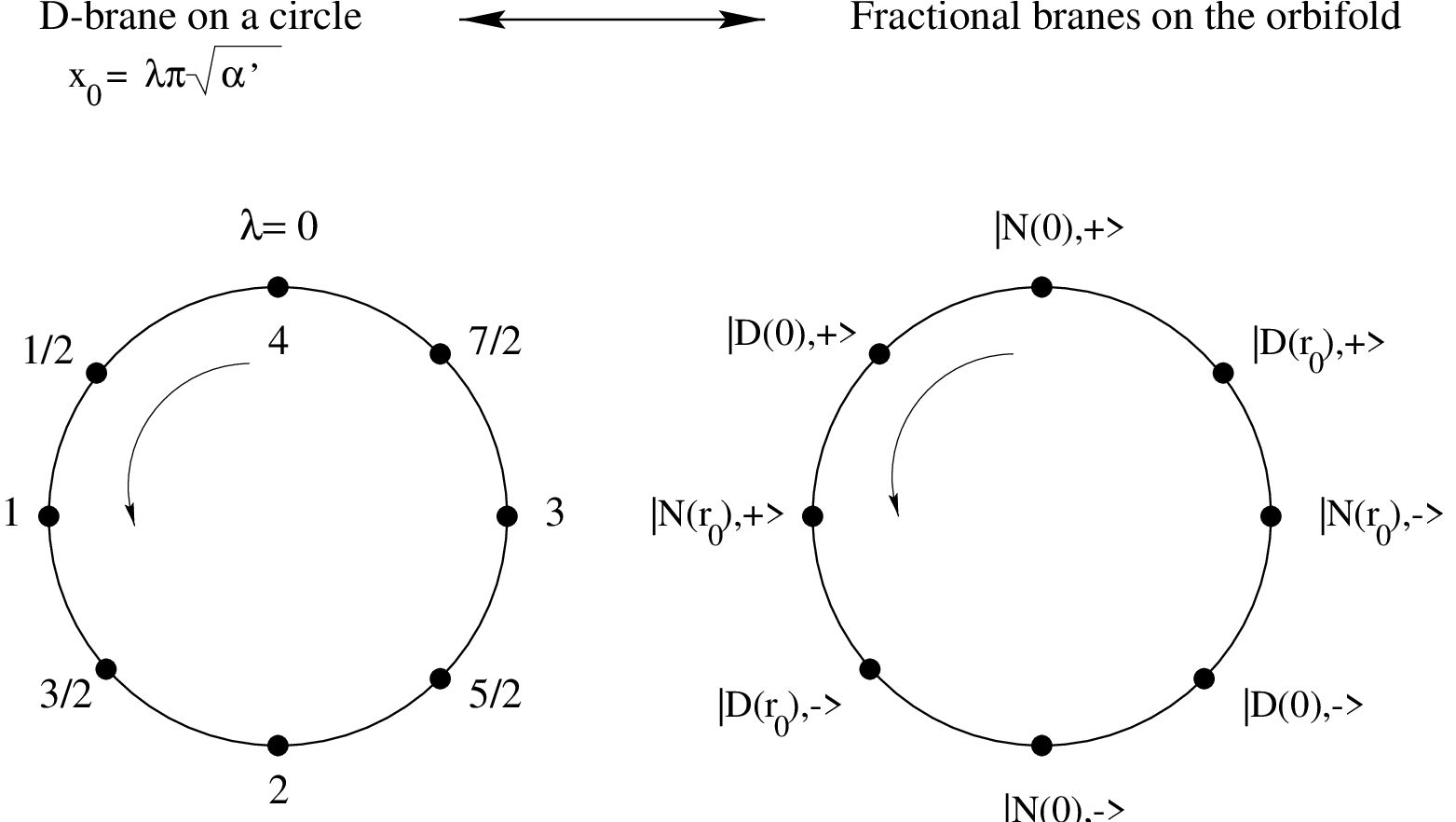}{10}{Comparison between the boundary states in
orbifold at self-dual radius and circle theory at twice self-dual radius.  On the right diagram, $0$ and $r_0$ refer to fixed
point values, while $\pm$ refer to the $Z_2$ representation each fractional brane realizes.}

\section{The Boundary State Interpretation}\label{bdystate}
Having described open string partition functions, the next task is to pull out interesting space-time physics in the closed string
channel.  We consider the calculation of the overlap between the deformed boundary state and various closed string states; these
correspond to one-point functions on the disk. Here we will be interested in extracting the overlap with lowest lying closed
string states, as they contain information about the center of mass position. In the untwisted sector, this is the tachyon, and
this problem has been studied extensively (\cite{SenReview} is a review). The overlap with the tachyonic vacuum in the infinite radius theory is
the function
\beq\label{fSen}
 f_\lambda (x) = \frac{1}{1+e^{ix}\sin(\pi \lambda)}
 + \frac{1}{1+e^{-ix}\sin(\pi \lambda)} -1 \ ,
\eeq
also corresponding to the open string disk partition function in the S-brane background.

One could attempt to get at these results by factorization of the annulus amplitudes in the closed string channel. The idea would
be that we can try to isolate the disk amplitudes via \beq \langle B;\tilde\lambda|\Delta|B;\lambda\rangle \to
\sum_\psi\langle\psi|B;\tilde\lambda\rangle^* \Delta_\psi \langle\psi|B;\lambda\rangle \eeq and we want to isolate
$\langle\psi|B;\lambda\rangle$ for suitable $\psi$. In the untwisted sector, we would like $|\psi\rangle$ to be a momentum $p$
tachyon. We have \beqn Z_1&=&\frac{1}{\sqrt{2}}\frac{1}{\eta(\tilde q^2)}\int_0^1 d\zeta_-\sum_{n\in\bZ}(\tilde
q^2)^{n^2/4}e^{2\pi i\alpha(\zeta_-)n}
 \eeqn
Because we are interested in information describing the center of mass positions, we need to isolate the contributions of Virasoro
primaries that do not correspond to oscillator excitations. This subset of states will build up $f_\lambda(x)$.

\subsection{Untwisted Sector: Self-Dual Radius}

At self-dual radius, we have $\zeta_-=0$, and $\alpha=(\lambda-\tilde\lambda)/2$, so the amplitude becomes
\beqn
Z_{1;sd}&=&\frac{1}{\sqrt{2}}\frac{1}{\eta(\tilde q^2)}\sum_{n\in\bZ}(\tilde q^2)^{n^2/4}e^{i \pi(\lambda-\tilde\lambda) n}
 \eeqn
It is tempting to simply take the above amplitude and discard the eta function $\eta (\tilde q^2)$, as this is usually associated
with oscillator contributions. We would obtain a phase $e^{i\pi\lambda n}$ for each $n$. However, this would not give the proper
$f_\lambda(x)$. The issue here is that the discrete primaries are also built out of the oscillators.  We should subtract the
contributions from both conformal descendants and the discrete primaries to identify the quantity $f_\lambda(x).$  This is in fact
what the $SU(2)$ formalism does for us -- it converts the annulus to the true Virasoro character, and gives a coefficient which is
related to an $SU(2)$ character, \beqn Z_1 &=&\frac{1}{\sqrt{2}}\sum_{j=0,1/2,1,\ldots}\chi^{Vir}_{j^2}(\tilde
q^2)\sum_{m=-j}^{j}\mathcal{D}^{(j)}_{m,m} \left(e^{2\pi i (\lambda-\tilde\lambda) J^1}\right)\nonumber \ . \eeqn This factorizes
into the Ishibashi states ${\cal D}_{-m,n}(e^{2\pi i\lambda J^1})|j,m,n\rangle\rangle$. The non-oscillator parts of this
correspond to $m,n = \pm j$ and we arrive at
\begin{eqnarray}
 \sum_p e^{ipx} \langle p,p|B;\lambda\rangle_{sd} &=&{\cal D}_{-j,j}(\cdot )\sum_p e^{ipx}
 \langle p,p|j,j,j\rangle\rangle+{\cal D}_{j,-j}(\cdot )\sum_p e^{ipx} \langle p,p|j,-j,-j\rangle\rangle\\ &&
 +{\cal D}_{-j,-j}(\cdot )\sum_p e^{ipx} \langle p,p|j,j,-j\rangle\rangle+{\cal D}_{j,j}(\cdot )\sum_p e^{ipx}
 \langle p,p|j,-j,j\rangle\rangle\nonumber\\
 &=& \sum_m e^{imx}(-\sin\pi\lambda)^{|m|}+\sum_m e^{im \tilde x}(i\cos\pi\lambda)^{|m|}
 \equiv f_\lambda(x)+\tilde f_\lambda( \tilde x) \ ,
\end{eqnarray}
where we reintroduced the variable $\tilde x$, T-dual of $x$. It is also possible to study the overlaps of boundary states with
low lying states within the dual theory on the circle of twice the self-dual radius. In this case however, there are some
subtleties involving the identification of zero modes in the two representations.

\subsection{Untwisted Sector: Infinite Radius}

At infinite radius, this computation simplifies: the $\bar f$ contribution decouples. We can see this directly given our infinite
radius expression \beq Z_1=\frac{1}{\sqrt{2}}\frac{1}{\eta(\tilde q^2)}\int_0^1 d\zeta_-\sum_{m\in\bZ}(\tilde q^2)^{m^2/4}e^{2\pi
i\alpha(\zeta_-)m},\eeq in the closed channel. Fortunately, once this is rewritten in terms of the Virasoro character, analogous
to the above discussion, the $\zeta_-$ integral is easily performed. The net effect is to reduce the result to
\begin{eqnarray}
 \sum_p e^{ipx} \langle p,p|B;\lambda\rangle_\infty &=&{\cal D}_{-j,j}(\cdot )\sum_p e^{ipx}
 \langle p,p|j,j,j\rangle\rangle+{\cal D}_{j,-j}(\cdot )\sum_p e^{ipx} \langle p,p|j,-j,-j\rangle\rangle\\
 &=& \sum_m e^{imx}(-\sin\pi\lambda)^{|m|} \nonumber \\
 &=& \frac{1}{1+e^{ix}\sin(\pi \lambda)}
 + \frac{1}{1+e^{-ix}\sin(\pi \lambda)} -1 = f_\lambda(x) \ ,
 \end{eqnarray}
so we have rederived (\ref{fSen}).

\subsection{Twisted Factorization: Self-Dual Radius}

Now in the twisted sector, we can follow the same path. Here, it is in fact easier,
because there is no subtlety
concerning the Virasoro character -- it is just what appears in the amplitude
\begin{equation}
Z_{g}=\frac{1}{\sqrt{2}}\frac{1}{\eta(\tilde q^2)}\sum_{\zeta_-}
\sum_{m\in\bZ}(\tilde q^2)^{(m-1/2)^2/4}e^{2\pi i\alpha(\zeta_-) (m-1/2)} \ .
\end{equation}
The non-oscillator part of this corresponds to the $m=0,1$ contributions.
Recall that there are no zero modes in the twisted sectors, so we just need to carefully enumerate the states that are contributing to the partition function. We note that since we have considered the most general deformation (with separate deformations $\lambda$,$\tilde\lambda$ on each boundary), we have enough information to do so.

At self-dual radius, we have only $\zeta_-=0$, and the amplitude reduces to
\begin{equation}
Z_{g}=\frac{1}{\sqrt{2}}\frac{1}{\eta(\tilde q^2)}\sum_{m\in\bZ}(\tilde q^2)^{(m-1/2)^2/4}
e^{i\pi (\lambda-\tilde\lambda) (m-1/2)} \ .
\end{equation}
Keeping only $m=0,1$, we find
\begin{equation}\label{eq:ztvac}
Z_{g,vac}=\frac{1}{\sqrt{2}}(\tilde q^2)^{1/16}
\left[ e^{i\pi (\lambda-\tilde\lambda)/2}+e^{-i\pi (\lambda-\tilde\lambda)/2}\right] \ .
\end{equation}
The ability to separate this amplitude into two factors depending on either $\lambda$ or $\tilde{\lambda}$ only, implies that
there are two orthogonal states contributing here, which we will denote by $|I\rangle_T$ and $|II\rangle_T$. We are finding that
\begin{equation}\label{Nlam}
|B,\lambda;0\rangle_{T;sd} = 2^{-1/4}e^{i\pi\lambda/2}|I\rangle_T
+ 2^{-1/4}e^{-i\pi\lambda/2}|II\rangle_T \ .
\end{equation}
We re-emphasize here that we are not considering the full boundary state, but really only its overlap with the twisted vacua. The
full boundary state contains oscillator excitations as well.  However, the dependence on the twisted vacua already contains the
information about the space-time positions.

At $\lambda=0$, this reduces to
\begin{equation}
|B,0;0\rangle_{T;sd} = 2^{-1/4} (|I\rangle_T
+ |II\rangle_T ) \equiv |N,0;0\rangle_{T;sd} \ ,
\end{equation}
which must coincide with one of the usual Neumann states
\cite{Oshikawa:1996dj}. We will verify this below.
Further, at $\lambda =1$, it reduces to
\begin{equation}
|B,1; 0\rangle_{T;sd} = i2^{-1/4}(|I\rangle_T
- |II\rangle_T) \equiv |N,1; 0\rangle_{T;sd} \ ,
\end{equation}
which must be the other Neumann state.
Note that these two states are orthogonal.
Next consider the Dirichlet states. We should get these by deforming to $\lambda=1/2$
and $\lambda=-1/2$. The former corresponds to a D0-brane
at the fixed point $X=0$, the latter to a D0-brane at $X=\pi\sqrt{\alpha'}$. So we identify\footnote{There is an unfortunate inconsistency in the literature which would seem to imply that $\lambda=1/2$ should correspond to $X=\pi\sqrt{\alpha'}$. However, those statements correspond to a situation with translational invariance (so a choice was made), whereas in the case of the orbifold, the position is fixed uniquely. This will be demonstrated carefully in the next subsection.}
\begin{equation} |B,1/2;0\rangle_{T;sd} = 2^{-1/4}(e^{+i\pi/4}|I\rangle_T+e^{-i\pi/4}|II\rangle_T)
\equiv |0\rangle_T \end{equation}
and
\begin{equation} |B,-1/2;0\rangle_{T;sd} = 2^{-1/4}(e^{-i\pi/4}|I\rangle_T+e^{+i\pi/4}|II\rangle_T)
\equiv |\pi\sqrt{\alpha'}\rangle_T \ .
\end{equation}
These states were called $|D(\phi_0)_T\rangle$ with $\phi_0=0,\pi r$ respectively
in \cite{Oshikawa:1996dj} ($r=\sqrt{\alpha'}$ for selfdual radius).
Note that they are orthogonal.
Now we can represent the states that we called $|I\rangle_T,|II\rangle_T$ in terms of the Dirichlet
states:
\begin{eqnarray}
|I\rangle_T &=& 2^{-3/4}(e^{-i\pi/4}|0\rangle_T
  + e^{i\pi/4}|\pi\sqrt{\alpha'}\rangle_T ) \nonumber \\
|II\rangle_T &=& 2^{-3/4}(e^{i\pi/4}|0\rangle_T
  + e^{-i\pi/4}|\pi\sqrt{\alpha'}\rangle_T ) \nonumber
\end{eqnarray}
and plug these back into the general expression for the deformed twisted boundary state (\ref{Nlam}).
We obtain
\begin{equation}\label{Nlam2}
|B,\lambda;0\rangle_T = \cos\left[\frac{\pi}{2}(\lambda -\half)\right] |0\rangle_T
+\cos\left[\frac{\pi}{2}(\lambda +\half)\right] |\pi\sqrt{\alpha'}\rangle_T \ .
\end{equation}
At $\lambda=0,1$ we then obtain the two Neumann boundary states
\begin{eqnarray}
|N,0;0\rangle &=& 2^{-1/2} (|0\rangle_T+|\pi\sqrt{\alpha'}\rangle_T) \nonumber \\
|N,1;0\rangle &=& 2^{-1/2} (|0\rangle_T-|\pi\sqrt{\alpha'}\rangle_T) \ .
\end{eqnarray}
These agree with the expressions in \cite{Oshikawa:1996dj}.
In fact, \cite{Oshikawa:1996dj} {\em assumed} the form of the two twisted Neumann boundary states.
Here we have {\em derived} them by deforming from known (Dirichlet) states.

\subsection{Twisted Factorization: Infinite Radius}

It is instructive to also look at this factorization in the infinite radius theory.
In this case, both $\zeta_-=0$ and $\zeta_-=1/2$ contribute and we arrive at
\begin{equation}
Z_{g}=\frac{1}{\sqrt{2}}\frac{1}{\eta(\tilde q^2)}\sum_{m\in\bZ}(\tilde q^2)^{(m-1/2)^2/4}
\left[e^{i\pi (\lambda-\tilde\lambda) (m-1/2)}+e^{i\pi (\lambda+\tilde\lambda-1) (m-1/2)}\right] \ .
\end{equation}
The contribution of the twisted vacua is
\begin{equation}\label{eq:ztvac}
Z_{g,vac}=\frac{1}{\sqrt{2}}(\tilde q^2)^{1/16}
\left[ e^{i\pi (\lambda-\tilde\lambda)/2}+e^{-i\pi (\lambda-\tilde\lambda)/2}
+e^{i\pi (\lambda+\tilde\lambda-1)/2}+e^{-i\pi (\lambda+\tilde\lambda-1)/2}\right] \ .
\end{equation}
We note that this expression factorizes
\begin{equation}\label{eq:ztvac}
Z_{g,vac}=\frac{1}{\sqrt{2}}(\tilde q^2)^{1/16}
\left[ e^{-i\pi \tilde\lambda/2}-ie^{i\pi \tilde\lambda/2}\right]
\left[ e^{i\pi \lambda/2}+ie^{-i\pi \lambda/2}\right]
\end{equation}
and thus we interpret this as only one state contributing,
\begin{equation} |B,\lambda\rangle_{T;\infty}=2^{-1/4}\left[ e^{i\pi \lambda/2}
+ie^{-i\pi \lambda/2}\right]|0\rangle_T \ .
\end{equation}
Note that the state $|B,-1/2\rangle_{T;\infty}$ is orthogonal to $|B,1/2\rangle_{T;\infty}$
as well as itself: that is, it decouples.
This corresponds to the fact that the second fixed point, present at finite radius,
has moved off to infinity, and the corresponding twisted boundary states decouple.
Similarly, by looking at integer $\lambda$, it is possible to see that $|\pi\sqrt{\alpha'}\rangle_T$
decouples. Thus, at infinite radius we only obtain the contribution from the twisted sector at
the remaining fixed point $X=0$.

\section{Back to the Lorentzian signature}
\label{Lorentzian}



After the Euclidean computations, we will now move back to Lorentzian
signature. As discussed in Section 2, in the case of the orbifold
there are some subtleties. We will also be interested in analyzing how the
brane decays into closed strings in the orbifold. We will compute the average
total energy and number densities of the emitted untwisted closed strings
and compare the calculation and the results with
those of \cite{Lambert:2003zr}.
The computations involve a prescription for a time integration contour,
which in turn is related to how the initial state of the brane is
prepared. A natural contour to use on the orbifold turns out to be the
Hartle-Hawking (HH) contour which was introduced in \cite{Lambert:2003zr}.
With this choice, on the fundamental domain we can interpret the unstable
brane to nucleate at the origin of time and then decay into closed strings.

Let us first review the standard case without the orbifold.
Upon Wick rotation back to the Lorentzian signature the overlap of the deformed
boundary state with the vacuum becomes
\begin{equation}
f(x^0) \equiv \mbox{}_{\infty}\bra 0|D(\pi \lambda )\ket_{\infty}
 = \frac{1}{1+e^{x^0}\sin (\pi \lambda)} +
   \frac{1}{1+e^{-x^0}\sin (\pi \lambda)} -1 \ .
\end{equation}
The full boundary state has the structure
\beqn
 |B\ket  = {\cal N}_p|B\ket_{X^0}|B\ket_{X}|B\ket_{bc}
\eeqn
where
\beqn
 |B\ket_{X^0} = f(x^0)|0\ket + \sigma (x^0)\alpha^0_{-1}\tilde{\alpha}^0_{-1}|0\ket + \cdots
\eeqn
with
\beqn
 \sigma (x^0)=\cos (2\pi\lambda )+1 -f(x^0) \ .
\eeqn
To compute the overlap with any on-shell closed string state, it is convenient to express
the vertex operators in the gauge
\beqn
 V = e^{iEX^0}V_{sp}
\eeqn
where $V_{sp}$ contains only the space part. The overlap then takes a simple form
\beqn
 \bra V|B\ket = \bra 0|e^{iEX^0}|B\ket_{X^0} \times {\rm (phase)}
\eeqn
yielding the amplitude
\beqn
  I(E) = i\int_{\Ccal} dx^0 e^{iEx^0} f(x^0)
\eeqn
with a suitable choice of the integration contour $\Ccal$.

In the case of the orbifold, the computations are simplest to perform in the covering space. As discussed in Section
\ref{orbifold}, a new feature is the need to double the Hilbert space by hand upon Wick rotating back to the Lorentzian signature.
This is because on the covering space a single quantum corresponds to two copies propagating into the opposite time directions
$X^0$ and $\Xtil^0 =-X^0$. The on-shell closed string states then take the form \beqn
  |V\ket = \frac{1}{\sqrt{2}} \left(\begin{array}{c}
                   e^{iEX^0}V_{sp}|0\ket \\
                   e^{iE\Xtil^0}V'_{sp}|0\ket
                 \end{array} \right) \ .
\eeqn
Similarly, the time direction part of the boundary state becomes
\beqn
  |B\ket_{X^0} = \frac{1}{\sqrt{2}}\left(\begin{array}{c}
                                           f(x_0)|0\ket \\
                                           f(x'_0)|0\ket
                                         \end{array}\right) + \cdots \ .
\eeqn
Note however that
\beqn
 f(x'^0)=f(x^0)= \frac{1}{1+e^{x^0}\sin (\pi \lambda)} +
   \frac{1}{1+e^{-x^0}\sin (\pi \lambda)} -1
\eeqn
because of time reflection symmetry.
The overlap with the closed string state becomes
\beqn
\label{VB}
 \bra V|B\ket = \half (e^{iEx^0}f(x^0)+e^{iEx'^0}f(x'^0)) \ .
\eeqn

Let us pause to compare the physical interpretation of the above with the standard full S-brane. The full S-brane corresponds to
formation and decay of an unstable brane, centered at the origin of the time axis. On the orbifold, the full Minkowski space is
replaced by the covering space, with a two-branched time direction. The unstable brane is centered at the origin of the time
coordinates, but decays into closed strings propagating into the opposite time directions, as illustrated in Figure 5.
\myfig{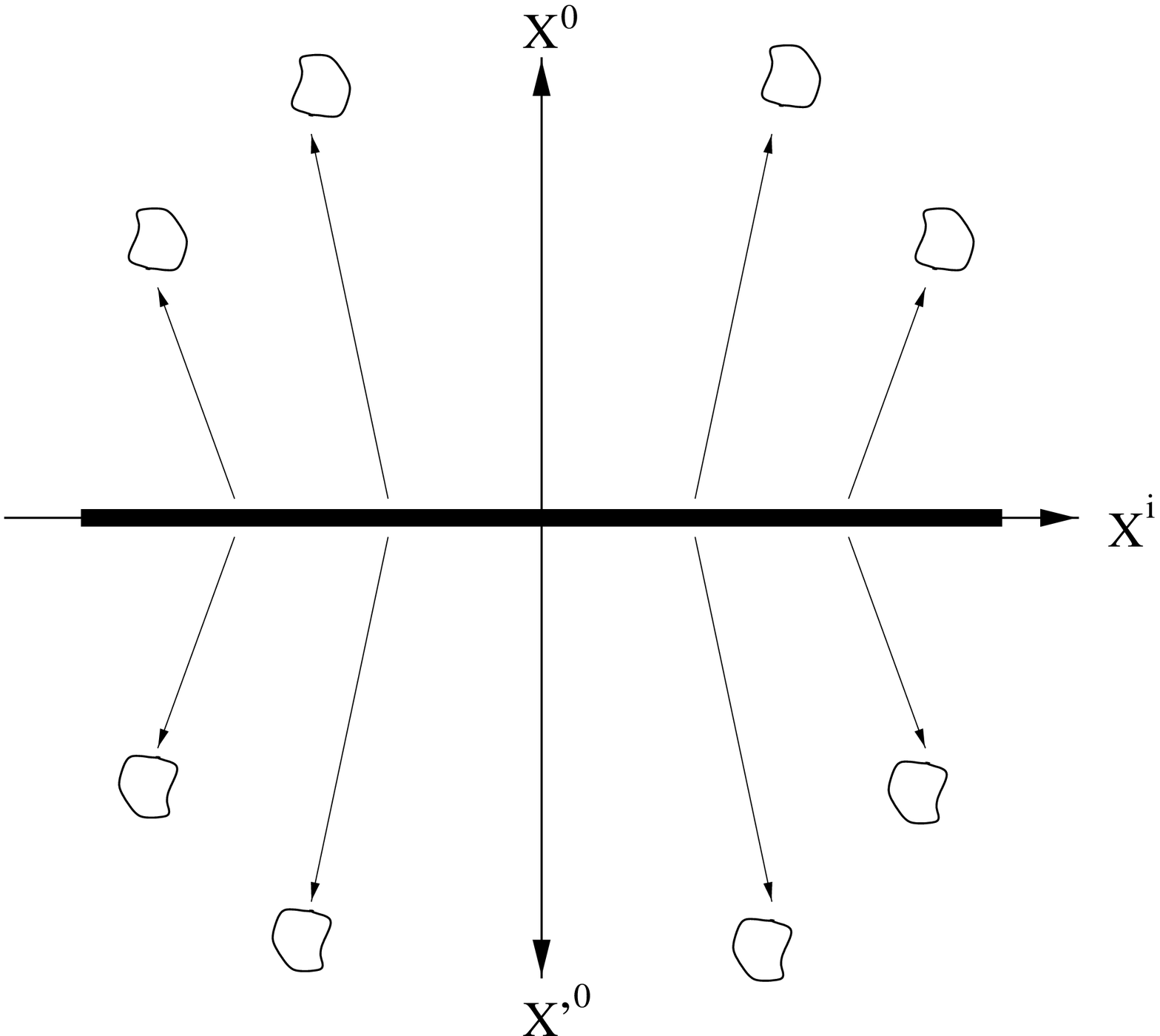}{5}{Untwisted closed string emission in the covering space.}

For the decay amplitude calculation, we then need a contour integration prescription. The fundamental domain has a semi-infinite
time axis, $X^0\geq 0$. We have set the decay of the brane to start at $X^0=0$. Since there is no past to $X^0=0$, we cannot build
up the brane from some closed string initial state. Instead, it is most natural to adopt the prescription in \cite{Lambert:2003zr}
for ``nucleating'' the brane via smeared D-instantons (see also \cite{Gaiotto:2003rm}) in imaginary time. This corresponds to
using a Hartle-Hawking time contour, coming in from $X^0=i\infty$ along the imaginary time axis to the origin and then proceeding
along the real time axis to $X^0=\infty$. For the actual calculation, we move back to the covering space where time runs from
$X^0=0$ to opposite time directions $X^0\rightarrow \infty$ and $X'^0=-X^0\rightarrow \infty$. The HH contour then maps to the
double contour with branches (see Fig. 6.) \myfig{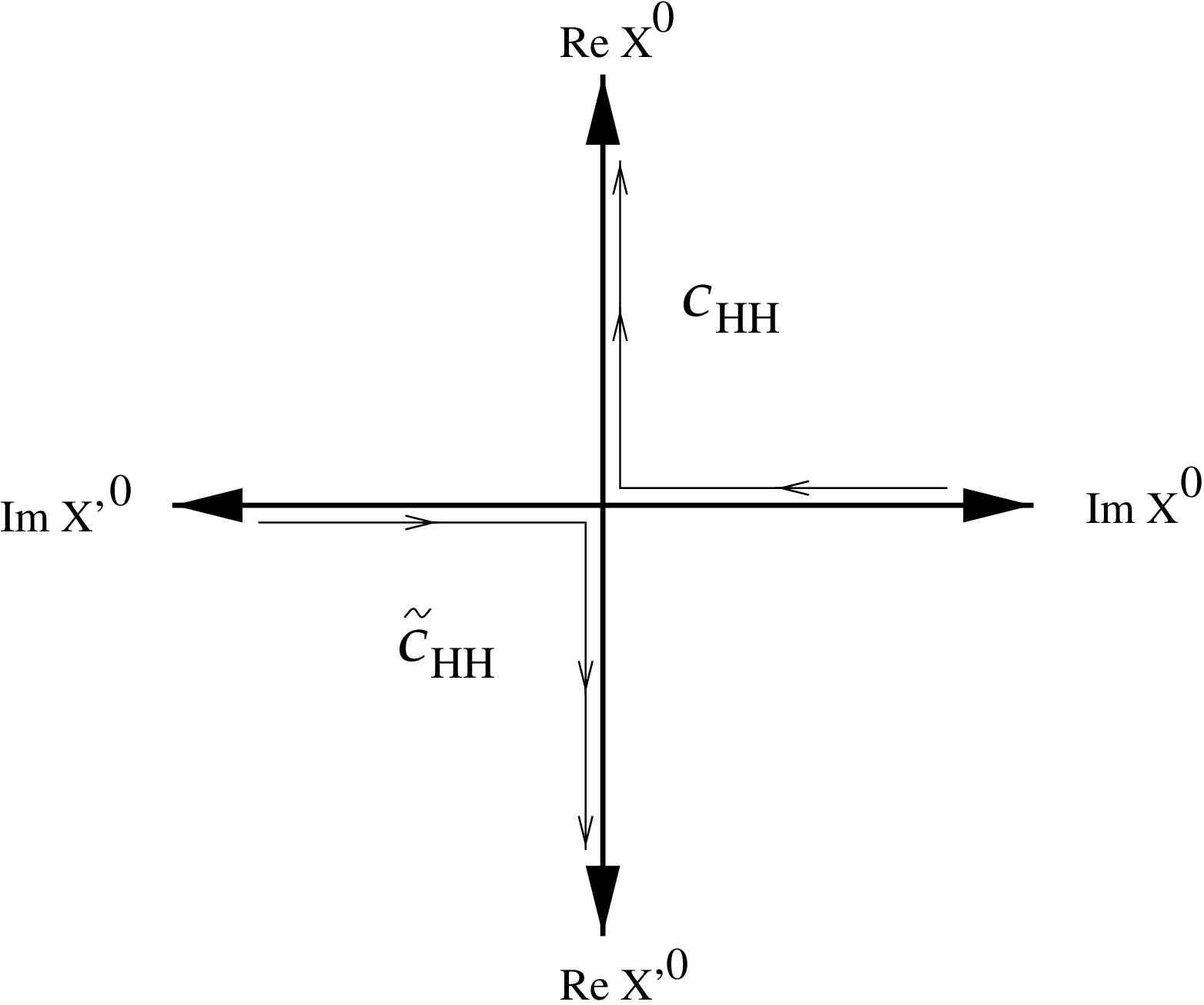}{7}{The double HH contour in the covering space.}
\begin{eqnarray}
 {\cal C}_{HH}, \tilde{{\cal C}}_{HH}:
 \ X^0,X'^0 = i\infty\rightarrow 0\rightarrow \infty
 \ .
\end{eqnarray}
Applying the contour to the overlap (\ref{VB}), we get
\begin{eqnarray}
 \bra V | B \ket &=& \half \int_{{\cal C}_{HH}}
e^{iEx^0}f(x^0)dx^0 +
 \half \int_{\tilde{\cal C}_{HH}} e^{iEx'^0}f(x'^0)
 dx'^0 \nonumber \\
 \mbox{} &=& e^{-iE\ln\lambda} \frac{\pi}{\sinh (\pi E)} \ ,
\end{eqnarray}
as in \cite{Lambert:2003zr} for the full brane with the HH
contour. The same is true for the total average energy and average
number densities for the produced untwisted closed strings on the fundamental
domain. The results are the same as in the standard case,
\begin{equation}
 \frac{\bar{N}}{V_p} \sim \sum_n n^{-1-p/4}
 \ ; \ \frac{\bar{E}}{V_p} \sim \sum_n n^{-1/2-p/4} \ ,
\end{equation}
where the sums are over the level numbers. To conclude, in
the orbifold the decay in the untwisted sector is quantitatively
the same as in the usual case.

The twisted sector is more problematic conceptually. Since the
twisted strings are localized in time, the concept of ``producing''
them in the decay is ill defined. Moreover, there are very
few physical states in the twisted sector. At present we
do not have much more to say about this matter.

\section{Conclusions and Outlook}\label{outlook}


The decay of unstable D-branes or D-brane configurations is an important open question in string theory. They may play a crucial
role in cosmology, and there already exist several scenarios making use of them. While spacetime orbifolds may be considered just
toy models, capturing some features of string theory in more general time-dependent backgrounds, they are nevertheless useful for
gaining insights into problems associated with quantum string theory. We have argued that the twisted sector which exists in
orbifolds may contribute to the decay of unstable branes. In particular, we have presented a detailed analysis of how to implement
the orbifold identifications into the decay, in a simple example, both in the open string and closed string formalism, and argued
that this leads to a new class of S-branes, which we have called fractional S-branes, in analogy to the fractional branes of
Euclidean orbifolds. We expect that the existence of fractional S-branes is a generic feature of spacetime orbifolds, and may
reflect some physics of D-brane decay in more general time-dependent backgrounds. They may also be relevant for the question of
resolution of spacelike singularities.

In particular, we have constructed a model where the D-brane decay has a semi-infinite duration, without
a prior build-up phase. This is in contrast to the full S-brane and half S-brane constructions, where
either the brane must first be formed from a fine-tuned closed string initial state, or the decay
starts from infinite past without any parameter to control its pace.
For potential applications of our construction, we can make at least the following speculative remarks.

(i) We have presented a model where an unstable brane, prepared at the ``Big Bang" origin of the spacetime, stores a large amount
of energy which then gets released in the decay into heavy closed string modes and the subsequent cascade into lighter
excitations. Presumably the large energy backreacts into the spacetime and converts it into an expanding cosmological model. The
initial condition, while defined at an initial spacelike singularity, is still under control because it has a well defined dual
formulation in terms of open string worldsheet theory. If the unstable brane is taken to be volume filling, it also provides a
homogeneous\footnote{Except for possible effects at the conical singularity.} initial condition. This may be compared with brane
cosmological models where a collision of almost parallel branes provides a homogeneous initial condition -- in our case the
homogeneity only involves a single brane.

(ii) The idea of an initial unstable brane at the Big Bang may be coupled
with string/brane gas cosmology \cite{Brandenberger:1988aj,Alexander:2000xv}.
Take the spacelike directions to be compactified to Planck Scale, and the initial unstable
brane to be wrapped in all directions. The brane then decays into closed strings, which interact
and presumably thermalize; thus brane decay could be viewed as the origin of the hot string gas.

(iii) In our orbifold construction, the covering space of the orbifold can be viewed as a model where two branches of spacetime
originate from the same Big Bang event. One may view the other branch and the images of closed strings in it simply as a
calculational trick, in analogy to the thermal ghosts in the real time formulation of finite temperature quantum field theory or
thermofield dynamics. But one could also view it as a real branch of the spacetime, so that the total spacetime contains a
multi-branched arrow of time.

(iv) If the other branch of the spacetime from the origin of time is viewed simply as a calculational
trick, one may ask if this trick could be applied in other cases. Consider for example the setup of
$D\bar{D}$-inflationary models, where a D-brane and an anti D-brane first approach each other, with
the scalar excitation from interpolating open string providing the rolling inflaton, and then form
an unstable system where the rolling tachyon provides an exit mechanism from inflation and may
be responsible for the subsequent reheating. It is not well understood how to actually model this
in the language of the open string sigma model. If the rolling tachyon has a hyperbolic cosine profile,
then it contains an unwanted stage where the tachyon rolls "up". If it only has an exponential profile,
then the decay starts at infinite past leaving no room for the inflationary stage. It may be possible
to develop a model, where the decay is modeled by our construction, viewing the other branch or time
direction from the origin simply as a calculational trick.

All the above comments are very tentative, but we believe they illustrate that there are exciting
possibilities ahead to unravel and study.

\bigskip

\noindent
{\Large \bf Acknowledgments}

\medskip

\noindent We thank V. Balasubramanian, N. Jokela, A. Naqvi for useful discussions and comments. In particular we thank P. Kraus
for discussions on boundary states and orbifolds at the early stages of this work. EKV thanks UCLAs IPAM and the organizers of the
Conformal Field Theory 2nd Reunion Conference, and the organizers of the Workshop on Gravitational Aspects of String Theory at
Fields Institute for hospitality while this work was in progress. SK would like to thank Matthias Gaberdiel, Andreas Recknagel,
and Masaki Oshikawa for discussions, and also the organizers of the July 2005 London Mathematical Society Durham Symposium on
"Geometry, Conformal Field Theory and String Theory."  EKV was in part supported by the Academy of Finland. SK was in part
supported by a JSPS fellowship. RGL and SN have support from the US Department of Energy under contract DE-FG02-91ER40709.

\bigskip

\providecommand{\href}[2]{#2}\begingroup\raggedright

\end{document}